# A Systematic Data Characteristic Understanding Framework towards Physical-Sensor Big Data Challenges

Zhipeng Ma[1,*], Bo Nørregaard Jørgensen[1,*], Zheng Grace Ma[1,*]

SDU Center for Energy Informatics, the Maersk Mc-Kinney Moller Institute, Faculty of Engineering, University of Southern Denmark Odense, Denmark

*Correspondence authors: zhma@mmmi.sdu.dk (Z.M.); bnj@mmmi.sdu.dk (B.N.J.); zma@mmmi.sdu.dk (Z.G.M.)

# Abstract

Big data present new opportunities for modern society while posing challenges for data scientists. Recent advancements in sensor networks and the widespread adoption of the Internet of Things (IoT) have led to the collection of physical-sensor data on an enormous scale. However, significant challenges arise in conducting high-quality data analytics within the realm of big data. To uncover big data challenges and enhance data quality, it is essential to quantitatively unveil data characteristics. Furthermore, the existing studies lack analysis of the specific time-related characteristics of physical-sensor data. Enhancing the efficiency and precision of data analytics through the big data lifecycle requires a comprehensive understanding of data characteristics to address the hidden big data challenges. To fill in the research gap, this paper proposes a systematic data characteristic framework based on a 6Vs model. The framework aims to unveil the data characteristics in terms of data volume, variety, velocity, veracity, value, and variability through a set of statistical indicators. This model improves the objectivity of data characteristic understanding by relying solely on data-driven indicators. The indicators related to time-related characteristics in physical-sensor data are also included for the analysis of temporal aspects in the physical-sensor data. Furthermore, the big data challenges are linked to each dimension of the 6Vs model to gain a quantitative understanding of the data challenges. Finally, a pipeline is developed to implement the proposed framework, and two case studies are conducted to illustrate the process of understanding the physical-sensor data characteristics and making recommendations for data preprocessing to address the big data challenges. The proposed framework is able to analyze the characteristics of all physical-sensor data, therefore, identifying potential challenges in subsequent analytics, and providing recommendations for data preprocessing. Furthermore, the characteristic indicators can be used to analyze other types of big data.

## 1. Introduction

Big data bring new opportunities to modern society including governments, industries, institutions, and so on [1], which hold great promise in uncovering subtle population patterns and heterogeneities that are not feasible with small-scale data. the International Data Corporation (IDC) has estimated that the amount of generated data will double every 2 years [2, 3]. In 2020, the daily data production totaled 175.9 exabytes, with an estimated increase to 495.9 exabytes by 2025 [4]. The immense data volume, diverse data content, and complex data structure introduce unique computational and statistical challenges in big data. These include issues related to scalability and storage bottleneck, noise accumulation, spurious correlation, incidental endogeneity, and measurement errors [5, 6]. As a result, it is essential to develop techniques for comprehending the characteristics of big data, which extract valuable insights and assist in addressing the challenges posed by big data, ultimately enhancing the reliability and accuracy of data analytics.

In real-world applications, big data often fail to yield valuable insights due to the aforementioned challenges. Coping with big data problems requires a lot of resources in addition to the direct adaption of existing analytics algorithms. In the realm of data analytics, raw data is seldom immediately suitable for processing. Instead, it typically undergoes multiple stages, including cleansing, integrating, and transforming, before progressing to further refinement, evaluation, and preparation as it moves through its lifecycle [7].

Enhancing the capacity to comprehend, manage, and make informed decisions based on vast volumes of data is a crucial concern in big data analytics, which aims to advance the selection of technologies to extract valuable insights from such substantial datasets for effective data processing. The huge amount of grown information hidden behind the datasets should be inferred and effectively managed to reveal the characteristics of big data, which should be the initial step before considering the application of any data pre-processing and analytic technologies. Data characteristics are models on the 'Vs' that describe the attribute and dimension information of data sources in detail [8, 9]. The number and the selection of vs vary in the different studies based on distinct requirements. Volume, variety, and velocity are the basic three Vs showing big data characteristics [10, 11]. It can be extended to 6 Vs by adding veracity, value, and variability [12-14]. In addition to the 6Vs framework, several other dimensions, such as validity [15], volatility [16], valence [15], and so on, have been developed in recent publications to encompass a broader and more comprehensive representation of data characteristics. In recent research, several new criteria have been introduced to broaden the understanding of data characteristics, such as visualization, validity, vitality, etc. [15-19]. These novel criteria are currently still in the development stage and potentially overlap the existing Vs concepts [20, 21]. Even Although the proposed Vs frameworks have provided precise and convincing definitions of all their dimensions, existing studies exhibit certain limitations when attempting to quantify these Vs to cover the comprehensive characteristics of large-scale data, primarily due to their sheer volume, multiple sources, and diverse formats [22].

The characteristics of big data not only furnish abundant information for methodology selection in data analytics but also unveil the challenges inherent in the field. The high dimensionality and large sample size raise unique challenges noise accumulation, computational cost, heterogeneity, and so on [1]. The data characteristics derived from Vs models are intricately linked to numerous challenges in many studies [10, 13, 23], and each 'V' represents one specific challenge. However, the limitations in quantifying Vs models result in non-quantitative representations of big data challenges. This may limit the provision of detailed and specific information regarding the challenges associated with certain datasets. It is necessary to utilize quantitative indicators to represent the big data challenges to make precise decisions on data preprocessing and data analytics.

The proposed Vs data characteristic frameworks are generally generic and can be employed to analyze datasets from different sectors. However, data derived from physical sensors have some specific characteristics and rely on the machine and system settings [22, 24]. With the availability of grand-scale use of measurement devices in multi-sensor systems [25], large-scale data are collected and managed in the Internet of Things (IoT), which involves intelligent digital sensors and networking technologies [26]. The errors of certain sensors or communication mechanisms in IoT systems may cause missing records and inconsistency of different variables, which damages the data quality and brings some challenges [27, 28]. Furthermore, a substantial portion of physical-sensor data is recorded as time series. The datasets are time-related tabular data with timestamps. The time-related characteristics, like timestamps, play vital roles in big data characteristics. Given the time-dependent nature of IoT systems, irregular time intervals or duplicate timestamps can lead to erroneous analyses. Current publications often lack time-related characteristic indicators, which should be integrated into the data characteristic framework for a more comprehensive understanding of physical-sensor data challenges.

To address the limitations mentioned above, this article proposes a systematic data characteristic understanding framework to analyze physical-sensor big data challenges. It aims to identify the characteristics of data recorded through physical sensors and understand potential challenges before adopting data preprocessing strategies. The data characteristic results provide crucial support for the development of data preprocessing pipelines. These pipelines are instrumental in enhancing the accuracy and trustworthiness of decision-making processes. Furthermore, the results foster increased transparency in data management, ensuring that information is handled more effectively. This framework provides a complete understanding of data characteristics with a 6Vs model. To reveal physical-sensor data characteristics quantitatively and objectively, each dimension of the 6Vs model includes a set of statistical indicators deriving information from raw data and several indicators are specifically for temporal data. Moreover, the evaluation metrics are developed for each dimension of Vs to quantitatively measure the overall quality of a dataset in each respective V-dimension. To understand the big data challenges systematically and make specific recommendations to cope with these challenges, each challenge is lined to one dimension of the 6Vs model and evaluated based on the statistical indicators. Furthermore, a pipeline is developed to implement the proposed data characteristics model, which includes timestamp understanding, value understanding, and feature understanding. Two case studies are conducted to demonstrate the process of data characteristic understanding. Two physical-sensor datasets are generated from the industrial processing sector and the transportation sector respectively. The results disclose the dataset characteristics and analyze the potential big data challenges, along with the corresponding preprocessing methods.

Reflecting the outlined scope of this article, its key contributions and novelty pertain to the following aspects:

- To address the limitations of quantifying the data characteristic model, this study develops a 6Vs model and each dimension involves a set of statistical indicators. This model improves the objectivity of data characteristic understanding by relying solely on data-driven indicators, evaluation metrics, and data-based information.

- To cope with the drawbacks of applying the general data characteristic model in physical-sensor big data, the study implements indicators related to temporal data into the data characteristic model to investigate time-related characteristics of physical-sensor data.

- To understand the challenges of physical-sensor big data systematically and quantitatively, each challenge is linked to one dimension of the 6Vs model.

- To visualize the application of the proposed framework, a pipeline to utilize the data characteristic understanding framework is developed in this article, and two case studies regarding physical-sensor data are conducted.

The remainder of this paper is arranged as follows. Section 2 introduces the background and the related work of the study. In Section 3, the proposed data characteristic understanding framework and the components are elaborated on and explained in detail. The pipeline to implement the proposed framework is presented in Section 4. Section 5 shows the two case studies as well as the results of data characteristic understanding. Furthermore, Section 6 discusses the data challenges in the two cases, compares the developed framework with existing models, and makes recommendations for data preprocessing. Finally, Section 7 concludes the study and presents the plan for future work.

## 2. Background and Related Work

### 2.1 Big data lifecycle

A vast volume of data is constantly being produced in the world, which is generated from diverse sources, such as The Internet of Things (IoT) [29], simulators [30], human activities [31], and so on. Large-scale and intricate datasets are defined as big data, which requires economical and advanced methodologies of digital information analytics for generating insight and supporting decision-making [5, 32, 33].[5, 32, 33]. The complexity of big data may result in a high false discovery rate in applications; in the meantime, challenges of big data analytics exist in all data processing phases, including data acquisition, data storage, and data analysis [20, 34]. Therefore, it is crucial to manage the data values across all phases with a global framework and elucidate the interconnections among these phases. The Data Lifecycle (DLC) model is proposed as an effective data management tool streamlining data organization and knowledge extraction within complex data systems, which highlights the sequential progression of data phases, outlines the management policies for each phase, and describes the relationship between these phases [6, 35, 36].

The Cross Industry Standard Process for Data Mining (CRISP-DM) is the de-facto standard and pipeline to develop data mining and knowledge discovery projects and has been widely employed in most data science projects [37, 38]. The original CRISP-DM model is illustrated in Figure 1. This model is a six-step cycle, and there exist iterative processes between 'business understanding' and 'data understanding', as well as between 'data preparation' and 'modeling'. The definition of the six steps is as follows[39, 40]:

(1) Business understanding: Clarify the project's objectives and define priority and success criteria;
(2) Data understanding: Gain a comprehensive understanding of the project's data and evaluate it as necessary;
(3) Data preparation: Execute essential data transformations, such as extracting target data, handling missing values, and dataset reconstruction;
(4) Modeling: Select an appropriate data processing model or framework;
(5) Evaluation: Utilize an application to assess the model's accuracy and versatility;
(6) Deployment: Summarize the process and share knowledge gained from the analysis.

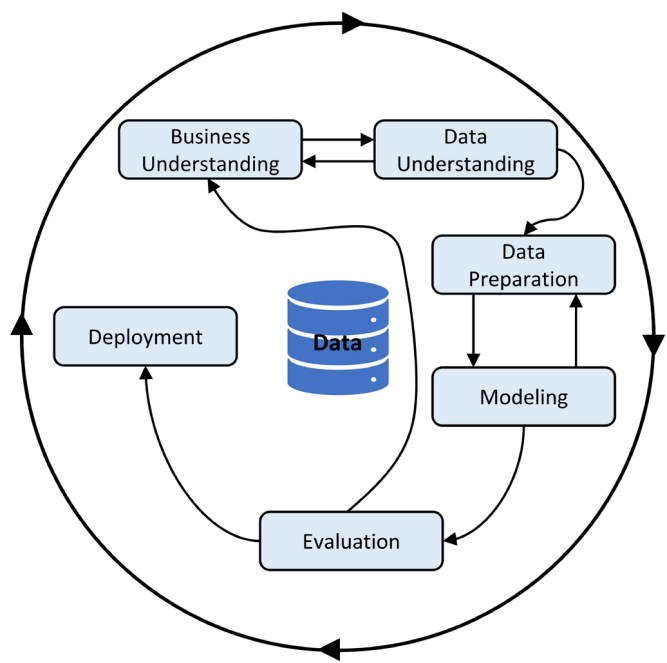

Figure 1. Original CRISP-DM process model [39].

In data science applications, The CRISP-DM model can be slightly adapted to many different domains, such as health care [41, 42], signal processing [43], sensor applications [40], production [44, 45], and so on. In the domain of physical-sensor big data processing, the automated preprocessing for data mining (APREP-DM) model has been proposed as a framework for sensor data analysis [40]. Physical-sensor data in this study are time-related tabular data with timestamps. This model replaces the data preparation step in the CRISP-DM model with the data preprocessing step, focusing on data cleaning and data transformation. Figure 2 shows the DLC model combining both APREP-DM and CRISP-DM frameworks, guiding the process of physical-sensor data science projects.

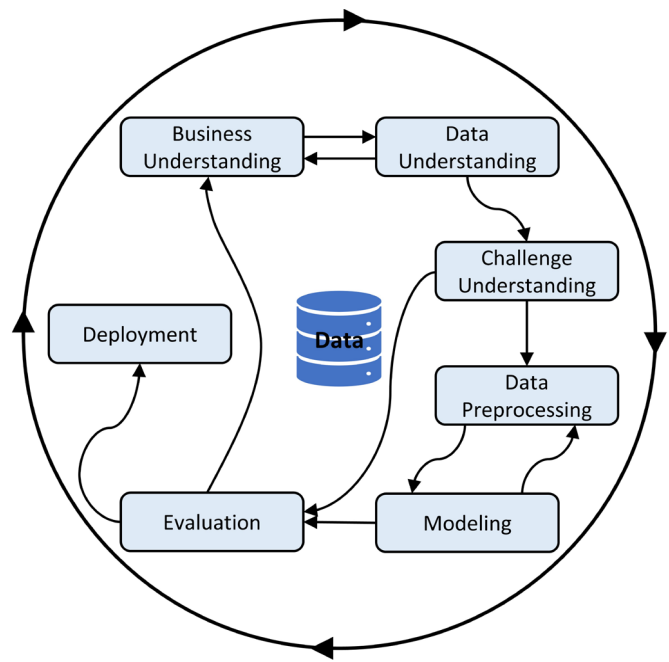

Figure 2. The DLC model for physical-sensor data.

The definition of the seven steps is introduced as follows:

(1) Business understanding: Clarify the project's objectives and define priority and success criteria;

(2) Data understanding: Gain a comprehensive understanding of the sensor-data characteristics in the project;

(3) Challenge understanding: Discover the challenges of physical-sensor data analytics based on data characteristics;

(4) Data preprocessing: Perform essential data preprocessing technologies, such as data cleaning and data transformations;

(5) Modeling: Select an appropriate data processing model or framework;

(6) Evaluation: Utilize applications to assess the data quality, and the model's accuracy and versatility;

(7) Deployment: Summarize the process and share knowledge gained from the analysis.

The framework in Figure 2 focuses on the understanding and data preprocessing steps, which are the most essential parts of data science projects [40]. The objectives and criteria in the business understanding step are used to determine the criteria of data characteristic understanding, and the results of data understanding are employed as feedback to optimize the business objectives and investigate the challenges of data preprocessing and modeling. Next, the data preprocessing strategies such as detecting outliers, handling missing data, and dimension reduction are selected based on the potential challenges. Subsequently, the developed model analyzes the preprocessed data, and the resulting insights offer valuable information for enhancing the data preprocessing. The evaluation step includes assessing the raw data quality based on the data understanding and challenge understanding and evaluating the model's accuracy and versatility. Finally, the results are concluded and discussed in the deployment step. This paper focuses on the data understanding and challenge understanding steps because they are essential tasks before data preprocessing. A systematic framework for understanding the characteristics and challenges of data empowers data scientists to gain valuable insights from raw data and make informed decisions regarding data preprocessing.

2.2 Big data characteristics in Vs models

The big data characteristics and challenges are defined as Vs, and the initial 3Vs are named Volume (large scale of data), Variety (various data formats and sources), and Velocity (rapid data generation and interaction) [10, 11]. Afterward, the initial 3Vs model, proposed to explain the technical and business implications of novel data management strategies, is extended to 6Vs to capture intricate big-data characteristics by incorporating veracity (trustworthiness and reliability of data), value (data with huge value) and variability (consistency in data formats, units, and scales) [13, 23, 46]. In recent research, many new criteria have been proposed to broaden the data characteristic understanding, such as visualization, validity, vitality, etc. [15-19] Table 1 outlines the different Vs in the analyzed literature. Such novel criteria are still in the development stage and potentially overlap the existing Vs concepts [20]; therefore, the 6Vs criteria are widely adopted in most literature to understand the big-data characteristics.

Table 1. Vs summarization in literature study.

| Ref. | Volume | Variety | Velocity | Veracity | Value | Variability | Visualization | Validity | Volatility | Valence | Vitality | Vincularity |
|---|---|---|---|---|---|---|---|---|---|---|---|---|
| [10] | ✓ | ✓ | ✓ | | | | | | | | | |
| [11] | ✓ | ✓ | ✓ | | | | | | | | | |
| [7]  | ✓ | ✓ | ✓ | ✓ | | | | | | | | |
| [9]  | ✓ | ✓ | ✓ | ✓ | ✓ | | | | | | | |
| [21] | ✓ | ✓ | ✓ | ✓ | ✓ | | | | | | | |
| [12] | ✓ | ✓ | ✓ | ✓ | ✓ | ✓ | | | | | | |
| [13] | ✓ | ✓ | ✓ | ✓ | ✓ | ✓ | | | | | | |
| [14] | ✓ | ✓ | ✓ | ✓ | ✓ | ✓ | | | | | | |
| [8]  | ✓ | ✓ | ✓ | ✓ | ✓ | | | | | | | ✓ |
| [15] | ✓ | ✓ | ✓ | ✓ | ✓ | | | ✓ | ✓ | ✓ | ✓ | ✓ |
| [17] | ✓ | ✓ | ✓ | ✓ | ✓ | ✓ | ✓ | | | | | |
| [18] | ✓ | ✓ | ✓ | ✓ | ✓ | ✓ | ✓ | | | | | |
| [16] | ✓ | ✓ | ✓ | ✓ | ✓ | | | ✓ | ✓ | | | |
| [19] | ✓ | ✓ | ✓ | ✓ | ✓ | ✓ | | | ✓ | | | |

Volume is the most visible big data characteristic, referring to the size and scale of data [47]. The evaluation of data volumes is relative, making it impractical to define a specific threshold for categorizing data volumes in big data analysis. Variety emphasizes the roles of structured, unstructured, and semi-structured data, which indicates the diversity of data forms in big data [19]. Velocity represents the speed at which data is received, stored, and processed, sometimes regarding real-time or near-real-time systems [6]. Veracity indicates the reliability and accuracy of data in terms of collection, processing methods, and trusted infrastructure [19]. In the context of scientific data, veracity has been partly defined as the aspect of data consistency that can be precisely characterized through statistical reliability [47]. Value emphasizes the significance of deriving actionable insights and value from big data. The concept of "low-value density" signifies that the value of big data is not strongly correlated with its volume; however, significant value can be achieved by processing large-volume data [6]. Variability denotes the inconsistency of data formats and structures due to multiple data sources with different characteristics [16]. The 6Vs framework shown in Figure 3 provides a holistic perspective on the big data characteristics with consensus definitions, helping navigate the challenges and opportunities that arise from managing and analyzing large and diverse datasets.

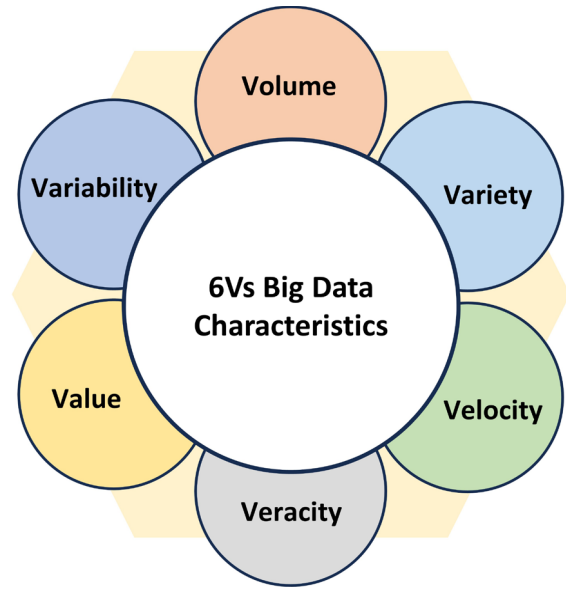

Figure 3. 6Vs big data characteristics.

Although The existing Vs models are capable of measuring big data characteristics in multiple dimensions, they still have limitations in comprehensively understanding big data. Firstly, the conventional 3Vs model falls short in covering the data attributes with the explosive growth of data size. However, the state-of-the-art Vs models involve several dimensions without mutual definitions, such as volatility and valence. Moreover, many developed Vs models only provide qualitative definitions or the boundary limit of each dimension. However, these models do not comprehensively reveal the data characteristics and the explanation of the boundary limits remains in a black box. Lastly, existing models concentrate on general big data characteristics and lack the incorporation of time-related indicators specific to physical-sensor data. Therefore, the proposed data characteristic understanding framework in this article utilizes a 6Vs model to measure physical-sensor data characteristics. This 6Vs model includes three conventional dimensions (Volume, Variety, and Velocity) and three novel and widely-used dimensions (Veracity, Value, and Variability) with mutual definitions in the literature study. Subsequently, each dimension is mapped with multiple statistical indicators to comprehensively understand the characteristics of physical-sensor big data. Some of these indicators involve time-related statistical metrics, such as timestamp-related measurements.

## 3. Data Characteristic Understanding Framework for Physical-sensor Big Data

### 3.1 Overview

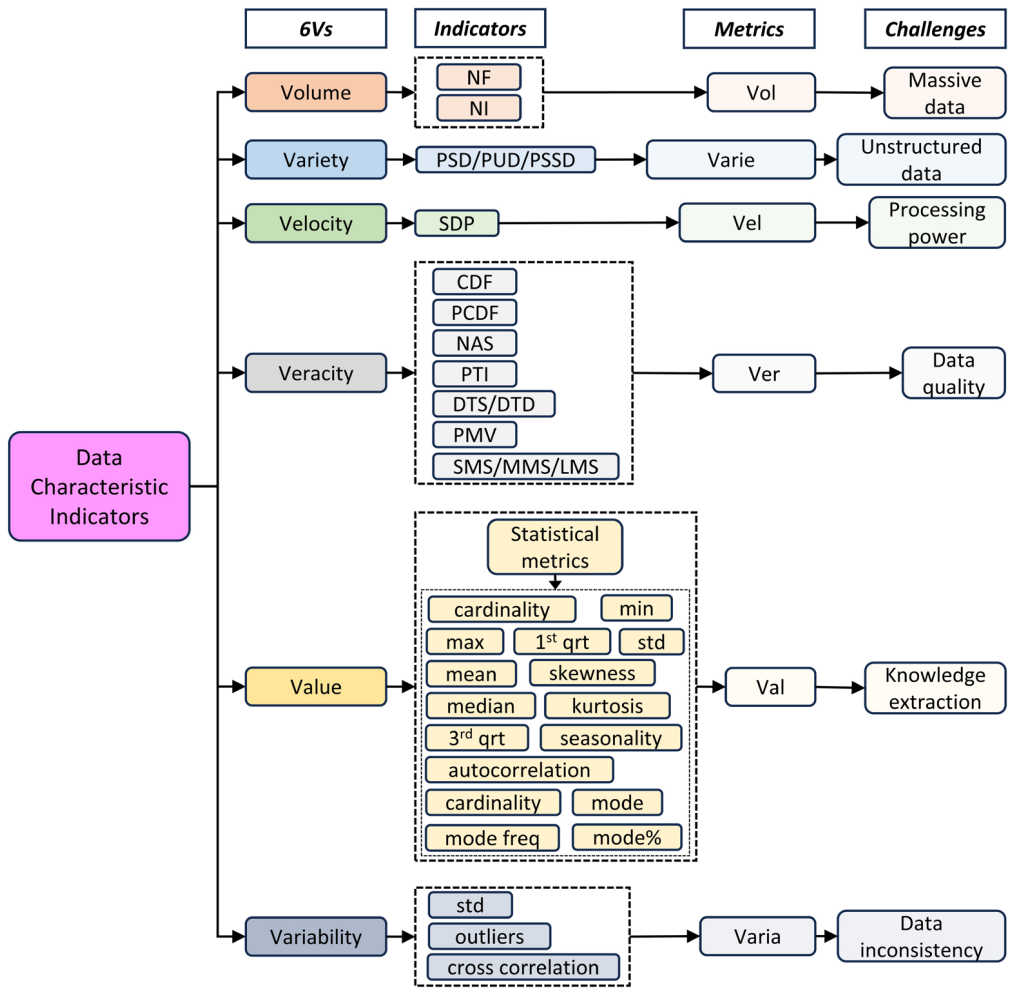

Figure 4. Data Characteristic Framework towards Data Challenges.

Figure 4 shows the general framework of data characteristics understanding for physical-sensor data and the corresponding big-data challenges. Physical-sensor data in this study are time-related tabular data with timestamps. The 6Vs model is selected as the base of this framework because it is a mature and comprehensive framework in data characteristic understanding. This 6Vs model includes volume, variety, velocity, veracity, value, and variability. Each of these six dimensions has clarified definitions in many studies [13, 23, 46] and does not overlap with others, ensuring distinct boundaries and precision within scholarly discourse. The models with less than six dimensions might lose several important data characteristics, whereas additional dimensions such as validity and volatility have no mature definitions and partly overlap with other dimensions.

In each dimension, some statistical indicators are selected or proposed to quantify the physical-sensor data characteristics. The objective of quantification is to provide precise and accurate measurements of each dimension of data characteristics. Moreover, quantification involves using standardized measurements, ensuring consistency and comparability across different studies, so that the proposed framework can be utilized in different datasets and to compare the results for better decision-making. The indicators in this study are specifically utilized for physical-sensor data analysis. Physical-sensor sensor data represents the data recorded by sensors to detect the parameters of real-world environments, machine operations, production qualities, and so on. The data of physical sensors are mainly recorded as temporal data and in numerical forms; therefore, the indicators regarding timestamps, probability distributions, and other relevant information are essential in data characteristic understanding. The details of the indicators, including the full name of the abbreviations, the definitions, and the calculation formulas, are described in Section 3.2.

The third part introduces the evaluation metrics for each dimension. Each metric integrates the general information from all indicators in the respective dimension to assess the overall quality of a dataset. The evaluation metrics indicate the quality of the dataset in the specific dimension, with the indicators providing detailed characteristics to substantiate the quality assessment. Therefore, to achieve a comprehensive understanding of data characteristics, it is essential to consider both the indicators and the metrics. The details of the metrics, including the definitions and the calculation formulas, are described in Section 3.2 following the presentation of the metrics in each dimension.

The fourth part of the proposed framework is the big data challenges. The challenges regarding physical-sensor data are identified through the analysis of data characteristic indicators, and based on the definition of each dimension of the 6Vs model. These challenges have a direct impact on data quality and provide valuable insights for effective data preprocessing. The details of the data challenges, involving the definition and the relationship with data characteristic indicators, are introduced in Section 3.3.

## 3.2 Data characteristic indicators and evaluation metrics

The data characteristic indicators are formulated based on the definition of the six dimensions outlined in the 6Vs model, aiming at quantifying the big data characteristics and providing precise and accurate insights on the physical-sensor data. These indicators serve as a valuable tool for analyzing the challenges posed by physical-sensor big data and shaping the methodologies for data preprocessing. Subsequently, an evaluation metric is defined in each dimension to quantitatively measure the overall quality of a dataset in each V-dimension.

### 3.2.1 Volume

Data volume corresponds to the size and scale of a dataset, and the number of features and instances (NF & NI) are two key indicators to represent the extent of the volume. A feature is a unique characteristic or attribute that provides valuable information about individual instances in a dataset [48, 49]. The uniqueness of features has a significant impact on the complexity and accuracy of data analysis, and a large number of features in the raw dataset provides an opportunity to choose informative, discriminating, and independent features that enable effective decision-making. An instance is a data point with multiple features in a dataset, and each instance has the same feature structure as the dataset [3]. The large instance size provides a reliable statistical representation of data distribution, allowing strong inferences on the data quality [50]. Moreover, more data instances can support to select of more informative and relevant features, which enhances the feature-selection model performance. However, it is impractical to define a universal threshold for big data volume, owing to the diverse influencing factors such as time, data type, and specific task requirements [6]. Different contexts necessitate distinct data volumes, reflecting the unique demands of various tasks.

The evaluation metric of the volume of a dataset $DS$ is defined as Equation 1 shows:

$$Vol(DS) = NF(DS) \times NI(DS) \tag{1}$$

where $Vol(DS)$ represents the score of the volume of the dataset $DS$, while $NF(DS)$ and $NI(DS)$ denote the two indicators NF and NI of the dataset $DS$.

A higher value of $Vol$ indicates a larger data volume, enabling the comparison of the data-processing task's demand to ascertain whether the data volume is adequate for the task.

3.2.2 Variety

Variety refers to the data forms in a dataset, including structured data, semi-structured data, and unstructured data [19]. Structured data is highly organized and easily decipherable by machine learning algorithms, stored in relational databases [3, 6]. In real-world applications, structured data involves sensor data, market data, consumption data, and so on [51]. Unstructured data are difficult to be analyzed by conventional data processing tools and methods, which typically include text and social media data [52, 53]. Semi-structured data does not follow the format of tabular data and cannot be stored in relational databases; however, it contains several structural elements such as tags and metadata in favor of data understanding [3, 6]. The most typical semi-structured data in physical-sensor systems are IoT data [51].

The distribution of data forms is the indicator to represent the variety characteristics as Table 2 demonstrates, including the percentage of structured data (PSD), percentage of unstructured data (PUD), and percentage of semi-structured data (PSSD). Each indicator represents the percentage of the volume of a particular form of data and the sum of these three indicators should equal one.

Table 2. Distribution of data forms.

| Percentage of structured data (PSD) | Percentage of unstructured data (PUD) | Percentage of semi-structured data (PSSD) |
|---|---|---|
| | | |

(PSD + PUD + PSSD = 1)

Equations 2-4 show the calculation of each indicator of data variety.

$$PSD = \frac{volume\ of\ structured\ data}{volume\ of\ all\ data} \times 100\% \qquad (2)$$

$$PUD = \frac{volume\ of\ unstructured\ data}{volume\ of\ all\ data} \times 100\% \qquad (3)$$

$$PSSD = \frac{volume\ of\ semi-structured\ data}{volume\ of\ all\ data} \times 100\% \qquad (4)$$

Equation 5 demonstrates the evaluation metric of the variety of a dataset $DS$.

$$Varie(DS) = \frac{PSD(DS)}{PUD(DS)+PSSD(DS)} \qquad (5)$$

where $Varie(DS)$ represents the score of the variety of the dataset $DS$, while the elements on the right side of the equation denote the data-variety indicators of the dataset $DS$.

This metric signifies the ratio between the structured data and non-structured data. A value greater than 1 suggests that a majority of the data is structured, while a value lower than 1 indicates that strategies should be implemented to convert most data into structured formats. If all data instances in the dataset are structured data, the $Varie(DS)$ is recorded as '+∞' to represent the best performance in data variety.

The classification of data forms can guide further data processing. While marching learning algorithms can be directly employed to analyze structured data, it is challenging to efficiently analyze unstructured and semi-structured data. This is due to the heterogeneous nature of the data sources, which present a variety of data types and representations. Therefore, it is necessary to convert them into structured data with data preprocessing techniques, including data cleaning, data integration, and data transformation [54-56].

3.2.3 Velocity

Velocity comprises the speed of data producing and data processing [3]. Big data does not always require immediate utilization; however, in certain domains, there exists a substantial advantage in acquiring up-to-the-second information and having the capability to respond accordingly. For example, devices in the cyber-physical system usually rely on the real-time operating system that enforces strict timing standards for execution [3]. As a result, they may encounter issues when data from a big data application fails to be delivered punctually. The speed of data producing (SDP) is utilized as the indicator of velocity because this study primarily focuses on data characteristics before processing. SDP denotes the frequency of data updated in the data storage platform such as IoT systems. In numerous domains such as cyber-physical systems and real-time applications, data is continuously generated and ingested by storage platforms. This constant stream of data presents unique demands on data storage, retrieval, and transmission systems, all of which need to keep pace with the SDP to ensure timely and accurate data processing.

Equation 6 defines the evaluation metric $Vel(DS)$ of the velocity of a dataset $DS$. Since only one indicator is defined in this dimension, it naturally serves as the metric for assessing data velocity.

$$Vel(DS) = SDP(DS) \qquad (6)$$

3.2.4 Veracity

Veracity represents the accuracy and reliability of big data, which is most relevant to the data quality and trustworthiness [3, 19]. Because of the increasing diversity of data sources and variety of big data, data can be incomplete, inconsistent, and noisy, negatively impacting big data analytics [1, 53]. In this study, several quantitative indicators, serving as measures of data quality and trustworthiness, are chosen or developed to illustrate data veracity in multiple dimensions.

The correctness of data formats (CDF) denotes the extent to which the data format of each feature matches the data processing requirements. For example, the correct data format of the feature "the number of the productions" should be integer. The CDF is recorded as "YES" or "NO", representing the correctness or not.

For multi-feature datasets, the percentage of the correctness of data formats (PCDF) can be utilized as the assessing criterion as Equation 7 shows.

$$PCDF = \frac{number\ of\ features\ with\ correct\ data\ format}{number\ of\ features} \times 100\% \tag{7}$$

The abnormal spikes represent the anomalous data points or subsequences in the time series, indicating the important abnormal events, such as production faults, system defects, and so on, which reflects the reliability of physical-sensor data collection [38, 50]. Data spikes are detected by rule-based strategies [57] or machine learning algorithms [58], and then the abnormal spikes are selected based on the real-world system properties. The number of abnormal spikes (NAS) is utilized to demonstrate the data reliability, and the happening time of the spikes as well as the spike values are also recorded for analyzing the abnormal spikes more comprehensively and systematically.

A time interval represents the amount of time between two given time points in the time series, and it should be constant in a stable system [24, 59]. However, owing to data communication delays in IoT systems or the presence of missing values, time intervals in real-world datasets may vary in different positions. This variability leads to inconveniences in time series analysis since many time series analysis strategies, such as first-order differencing and sliding windows, assume a fixed time interval. In this research, the percentage of normal time intervals (PTI) serves as the chosen evaluation criterion as Equation 8 displays. A normal time interval is defined by the system settings.

$$PTI = \frac{number\ of\ normal\ time\ intervals}{number\ of\ time\ intervals} \times 100\% \tag{8}$$

In a time series, each timestamp should record only one value, whether in a single dimension or multiple dimensions. Duplicate timestamps indicate that multiple values are associated with the same timestamp, typically as a result of recording errors. Therefore, duplicate timestamps should be recorded as abnormal data. The duplicate timestamps are classified into two categories: duplicate timestamps with the same value (DTS) and duplicate timestamps with different values (DTD). In further processing, DTS can be easily integrated into one data instance by removing repeated records, whereas in DTD some advanced methods should be developed to find out the most trustworthy value among the records.

The presence of missing values degrades the data quality and adds complication and uncertainty to big data processing [1, 60]. The missing data are categorized into three types, including missing completely at random (MCAR), missing at random (MAR), and missing not at random (MNAR) [60, 61]. The majority of the missing physical-sensor data are assumed to be MAR, where there exist differences between missing and observed values but these differences can be explained by other observed data [62, 63]. The percentage of missing values for each feature (PMV) can be utilized to assess the data complementation as Equation 9 shows.

$$PMV = \frac{number\ of\ missing\ values\ in\ a\ certain\ feature}{number\ of\ values\ in\ this\ feature} \times 100\% \tag{9}$$

Except for the PMV, the missing time span is another important index demonstrating the information of missing values, which indicates a time span during which all values are missing. Different length of time span requires distinct imputation methods. For example, in the short span, some statistical values based on the distribution such as mean, median, and mode [60] can be used to impute the missing values. However, in the long span, the data distribution is unclear and the advanced deep-learning algorithms should be used to estimate the missing values [62]. In this study, the missing time spans are classified into three groups based on their length, including short-term missing span (SMS), medium-term missing span (MMS), and long-term missing span (LMS). The classification threshold is determined by the application requirements and constraints.

Equation 10 demonstrates the evaluation metric $Ver(DS)$ of the veracity of a dataset $DS$.

$$Ver(DS) = (1 - PCDF(DS)) \times W_{41} + \frac{NAS(DS)}{NI(DS)-2} \times W_{42} + (1 - PTI(DS)) \times W_{43} + PMV(DS) \times W_{44} \tag{10}$$

$Ver(DS)$ consists of four components associated with the weights $W_{41}$ - $W_{44}$ respectively. Each component is represented by a percentage value derived from the veracity indicators with values ranging from 0 to 1. The indicators 'NAS' and 'PMV' are computed for each feature, and the average values of these two indicators across all features are utilized as the values in Equation 10. The values of all five weights are set at $\frac{1}{4}$ by default and can be further adjusted according to the application requirements. The sum of the four weight values should equal 1. Therefore, the value of $Ver$ ranges from 0 to 1, and a higher value indicates lower veracity.

3.2.5 Value

Value represents the context and usefulness of data for deriving insights and facilitating informed decision-making, which is the ultimate goal of working with data [19, 35]. It is essential to identify how data can contribute to the attainment of application objectives and create a beneficial impact. Deriving value from big data is unique in different projects and is intricately related to the techniques employed for data processing. To Investigate the potential value of big data, the statistical metrics about data distribution can be calculated and analyzed. Table 3 lists the definitions of the 15 factors of data values.

Table 3. Definitions of the factors of data values [32, 64-66].

| Factors | Definitions |
|---|---|
| cardinality | the number of unique values |
| min | the minimum value of the data |
| 1st qrt | the value located at the first quarter of the values in ascending order |
| mean | the average value of the data |
| median | the median value of the data |
| 3rd qrt | the value located at the third quarter of the values in ascending order |
| max | the maximum value of the data |
| std | the standard deviation of the data |
| skewness | Skewness is a measure of the asymmetry of the probability distribution of a real-valued random variable about its mean. The negative skew indicates that the tail is on the left side of the distribution, and the positive skew indicates that the tail is on the right. |
| kurtosis | Kurtosis is a measure of the "tailedness" of the probability distribution of a real-valued random variable. The kurtosis for a standard normal distribution is 3. Kurtosis more than 3 indicates a "peaked" distribution and kurtosis less than 3 indicates a "flat" distribution. |
| seasonality | repeating cycle in the series with fixed frequencies (hour of the day, week, month, year, etc.). A seasonal pattern exists in a fixed known period. |
| autocorrelation | the correlation of a signal with a delayed copy of itself |
| mode | the most frequent value in the feature |
| mode freq | the frequency of the mode value in the feature |
| %mode | the percentage of the mode value |

Continuous features and categorical features are two distinct types of variables. Continuous features consist of numeric variables with an infinite number of values between any two data points, whereas categorical features represent characteristics that can be divided into different categories without any inherent order or priority. Consequently, due to their distinct data properties, the factors revealing their data values are also different. Table 4 and Table 5 show the statistical data characteristics of a continuous feature and a categorical feature respectively. The factors associated with continuous features reveal the characteristics of a distribution. However, unlike continuous data, categorical data does not adhere to a mathematical distribution of values. Therefore, the factors revealing the percentage and distribution of each category are employed.

Table 4. Statistical data characteristic factors of a continuous feature.

| cardinality | min | 1st qrt | mean | median | 3rd qrt | max | std | skewness | kurtosis | seasonality | autocorrelation |
|---|---|---|---|---|---|---|---|---|---|---|---|
| | | | | | | | | | | | |

Table 5. Statistical data characteristic factors of a categorical feature.

| cardinality | mode | mode freq | mode% | seasonality | autocorrelation |
|---|---|---|---|---|---|
| | | | | | |

Equation 11 displays the evaluation metric $Var(DS)$ of the value of a dataset $DS$.

$$Var(DS) = \frac{number\ of\ invalid\ indicators}{number\ of\ all\ potential\ indicators} \tag{11}$$

$Var(DS)$ signifies the percentage of the invalid indicators regarding data values, whose value ranges from 0 to 1. A higher $Var$ value indicates a less comprehensive data value representation.

3.2.6 Variability

Variability signifies the inconsistency in data flows, measuring the spread or dispersion of data values and providing insights into the extent to which individual data points diverge from the central or mean value [16]. Data patterns and characteristics are subject to change over time or in response to different scenarios and operations. Variability provides users with a tool to quantify the extent of variation in datasets, thereby aiding in the assessment of the uncertainty or unpredictability in a dataset. The standard deviation (std) in Section 3.2.5 quantifies the dispersion of values and can thus serve as an index of variability.

An outlier is a data instance that significantly differs from the other observations, representing the abnormal variability of the dataset [67, 68]. Outliers can be detected by the IQR rule and graphically illustrated by the boxplot. Equations 12-14 outline the formulas to identify outliers.

$$upper\ boundary = Q3 + 1.5 \times IQR \tag{12}$$

where Q3 represents the 75th percentile of the data and IQR is the interquartile range shown in equation 14.

$$lower\ boundary = Q1 - 1.5 \times IQR \tag{13}$$

Where Q1 denotes the 25th percentile of the data.

$$IQR = Q3 - Q1 \tag{14}$$

The values larger than the upper boundary or smaller than the lower boundary are defined as outliers. In practical applications, the quartile values Q1 and Q3, as well as the coefficient '1.5,' can be adjusted to suit specific requirements. In physical-sensor systems, it is advisable to incorporate additional constraints, which may be derived from established rules of thumb or system settings, in order to effectively limit outlier detections.

$$outlier\ rate = \frac{number\ of\ outliers\ in\ a\ certain\ feature}{number\ of\ values\ in\ this\ feature} \tag{15}$$

The outlier rate in Equation 15 is calculated as an indicator to represent the overall outliers in each feature.

Correlation measures the variability between features, referring to the degree to which a pair of features are related [66, 69]. Cross correlation is an established and reliable correlation measure between two univariate time series by considering the time delay. For two time series $X = \{X_1, ..., X_n\}$ and $Y = \{Y_1, ..., Y_n\}$, the cross correlation with time delay $k$ $R_k(X, Y)$ is

$$R_k(X, Y) = \frac{\sum(X_t - \bar{X})(Y_t - \bar{Y})}{\sqrt{\sum(X_t - \bar{X})^2}\sqrt{\sum(Y_t - \bar{Y})^2}} \tag{16}$$

where $\bar{X}, \bar{Y}$ denote the mean of $X$ and $Y$ respectively.

Equation 17 demonstrates the evaluation metric $Varia(DS)$ of the variability of a dataset $DS$.

$$Varia(DS) = Nstd(DS) \times W_{51} + PO(DS) \times W_{52} + (1 - VC(DS)) \times W_{53} \tag{17}$$

where $Varia(DS)$ denotes the score of the variability of $DS$, $Nstd(DS)$ represents the normalized standard deviation defined in Equation 18, $PO(DS)$ stands for the percentage of outliers defined in Equation 19, and $VC(DS)$ signifies the validity of cross correlation defined in Equation 20.

The indicator 'std' represents the standard deviations of each feature in a dataset. In Equation 18, to calculate the $Nstd$ and ensure that its value falls in the range of [0, 1], the 'std' values are transformed to the range [0, 1] via min-max scaling. Subsequently, the average value of the transformed 'std' values are computed to obtain $Nstd$.

$$Nstd(DS) = Ave(Scaling(std)) \tag{18}$$

The 'outlier rate' denotes the percentage of outliers in a feature as Equation 15 shows. The $PO$ in Equation 19 computes the average of all values regarding 'outlier rate' in the dataset to show the dataset-level percentage of outliers.

$$PO(DS) = Ave(outlier\ rate) \tag{19}$$

The $VC$ stands for the validity of cross correlation, indicating the percentage of high correlation values. The high correlation value is defined as the value larger than 0.7 in this paper.

$$VC(DS) = \frac{number\ of\ high\ correlation\ values}{number\ of\ correlation\ values} \tag{20}$$

$Varia(DS)$ is composed of three components associated with the weights $W_{51}$ - $W_{53}$ respectively. Each component is represented by a percentage value derived from the veracity indicators with values ranging from 0 to 1. The values of all three weights are set at $\frac{1}{3}$ by default and can be further adjusted according to the application requirements. The sum of the three weight values should equal 1. Therefore, the value of $Varia$ ranges from 0 to 1, and a higher value indicates high data variability and high data inconsistency.

3.3 Big data challenges

The big data challenges are generated through a literature study, and then mapped to distinct dimensions in the 6Vs model based on their definitions. The objective is to propose a framework that guides future research in addressing these challenges effectively. The data preprocessing recommendations to address the challenges are formulated based on the characteristic indicators outlined in Section 3.2.

3.3.1 Volume

The challenge in data volume is to develop adequate technologies to effectively handle massive data [70]. Managing, storing, and processing datasets with a substantial number of instances and features poses significant difficulties. For instance, large-scale datasets might not be processed rapidly due to the limited computational ability of processors. Moreover, high-dimensional data have the "curse of dimensionality" due to the sparse data distribution, less reliable nearest neighbors, and so on. The characteristic indicators of data volume are NF and NI, representing the number of features and instances. This dimension serves as the general representation of the size of a dataset. Consequently, the methods to mitigate the negative impacts of the large data size, such as feature selection, dimensionality reduction, and data splitting, should be chosen by combining the information generated from other data characteristic dimensions.

3.3.2 Variety

Handling data heterogeneity is a challenge in data variety because traditional tools like SQL do not perform well in storing semi-structured or unstructured data [71]. In physical-sensor data, there exists some semi-structured and unstructured data. Semi-structured physical-sensor data are text-based representations of structured data and stored as key-value pairs and ordered lists, such as JSON and XML data. Unstructured physical sensor data involves log files in the IoT devices. As most popular AI algorithms only perform well in processing structured data, it is essential to transform all unstructured data and semi-structured data into structured data. ML and AI technologies can be employed to extract relevant information from unstructured data and map the extracted information to the defined schema for the structured data. For example, the information in the text data is extracted through natural language processing methods, and the necessary entities such as dates, names, and locations are mapped to a defined structure. The semi-structured data are easier to transform, as specific transformation tools have been developed and can be directly utilized to transform the semi-structured data into the desired structured format.

### 3.3.3 Velocity

The traditional algorithms and systems cannot provide sufficient data processing power while the business partners require real-time response and in-time predictive maintenance of the system [72]. The scope of this study does not encompass data processing velocity, as the primary focus is on the examination of raw data before undergoing preprocessing. Data velocity in this study represents SDP (speed of data producing), denoting the frequency of data updated in the data storage platform. Data should be updated in a short time interval to avoid significant information loss; however, the hardware might pose a challenge in achieving the required data update velocity. The challenge in data update velocity stems from the time required for sensors to record data, coupled with the inherent time delay in communication between the sensors and the storage platform. Therefore, high-quality sensors and communication devices should be implemented in the data processing systems to improve the velocity.

### 3.3.4 Veracity

The challenges associated with data veracity highlight the issues of big data quality, including incorrect data format, abnormal spikes, inconsistent time intervals, duplicate timestamps, and missing values. Despite large-scale data being available for utilization, low-quality data could result in inaccurate and unreliable decision-making [73].

Incorrect data formats can lead to computing errors and system failures. For instance, recording an integer value as a string can result in calculation failures since a string lacks the inherent capability to participate in mathematical operations. Therefore, all data formats should undergo assessment, and any formats found to be improper should be corrected.

The abnormal spikes represent the anomalous data points or subsequences in the time series, indicating the important abnormal events detected by sensors. The abnormal spikes can be detected by rule-based strategies [51] or machine learning algorithms [52]. Next, the extracted abnormal spikes can be analyzed to identify potential system faults, enhancing the overall diagnostic capabilities. Moreover, the values of the abnormal spikes should be replaced with reasonable alternatives to improve the reliability of the analysis conducted on the entire sequence. The missing data imputation technologies introduced at the end of this section can be utilized to estimate the alternatives.

Due to data communication delays in IoT systems or the occurrence of missing values, time intervals in physical-sensor datasets may exhibit variations in different positions, despite the default settings requiring constant time intervals. This variability results in challenges in time series analysis since several time series analysis strategies, such as first-order differencing and sliding windows, assume a fixed time interval. The irregular time intervals can be adjusted by considering the underlying reasons for the inconsistency. If the inconsistency is attributed to time delays, the timestamps can be directly adjusted according to the default settings. In cases where missing values contribute to the variability, employing missing data imputation technologies can effectively address this challenge.

The duplicate timestamps are the result of recording errors, and they are categorized into DTS (duplicate timestamps with the same value) and DTD (duplicate timestamps with different values) in this study. In further processing, DTS can be straightforwardly integrated into one data instance by removing repeated records, whereas in DTD some advanced methods, including statistical analysis and AI algorithms, should be developed to determine the most reliable value among the records.

The prevalence of missing data not only diminishes performance in monitoring decisions but also influences the immediate utilization of data analytics applications that rely on reliable access to accurate data in the system. The most straightforward strategy to deal with missing values is to delete the corresponding data points. However, this method results in inconsistent time intervals in time series analysis, as the removal of values leads to gaps in the temporal sequence, affecting the overall integrity of the analysis. As an alternative solution, one can select a continuous subsequence without any missing values for further analysis. This method is effective for time series data with a small number of missing values, particularly when they are concentrated within limited periods. However, when numerous missing values are distributed throughout the dataset, the selected subset may be of a small size and may not adequately capture the characteristics of the entire dataset. In such scenarios, missing data imputation methods, such as regression models [74] and GANs [75] can be employed to estimate and fill in the gaps in the dataset.

### 3.3.5 Value

The challenge in the data value dimension lies in the inability to extract useful information from numerous data sources and uncover the hidden value within the recorded data [70, 73]. The data characteristics introduced in Section 3.2.5 are statistical metrics pertaining to data distribution. To extract valuable insights from the data, it is imperative to conduct further analysis of the statistical values. The preprocessing steps, including data cleaning and transformation, heavily depend on the information derived from the data distribution. For instance, the decisions for dealing with missing values rely on a thorough understanding of the distribution of gaps in the sequence. Therefore, the computed statistical values ought to be connected with calculations in other characteristic dimensions to obtain a more comprehensive and accurate understanding of the collected physical-sensor big data. This approach contributes to making more reliable recommendations for data preprocessing.

### 3.3.6 Variability

Data variability signifies the inconsistency of the data points and features by measuring the outliers and cross-correlation between features. Outliers have the potential to introduce noise during model training, thereby distorting the outcomes of data analytics. Additionally, interpreting cross-correlation results in time series analysis is challenging due to the presence of time lags. To mitigate bias in datasets and uncover variables linked to target features, it is imperative to employ various strategies addressing the challenges posed by data inconsistency.

Detected outliers should be substituted with reasonable values, and models designed for imputing missing values can be utilized to estimate the data points. Furthermore, it is crucial to analyze the factors contributing to the outliers, such as potential sensor recording errors or operational faults.

Cross-correlation stands out as an easy-implemented method for examining the relationship between features utilizing a coefficient. However, for a more comprehensive analysis of feature relationships, it is advisable to incorporate additional techniques such as feature engineering [76] and causal inference [77] in the data preprocessing stage.

## 4. Pipeline of data characteristic understanding

Figure 5 demonstrates the data characteristic understanding pipeline. Data characteristic understanding in this pipeline includes timestamp understanding, value understanding, and feature understanding, which are shown in three dashed rectangle blocks, in which the light-blue blocks represent the calculation of data characteristic indicators, and the light-orange blocks represent the necessary data adjustment.

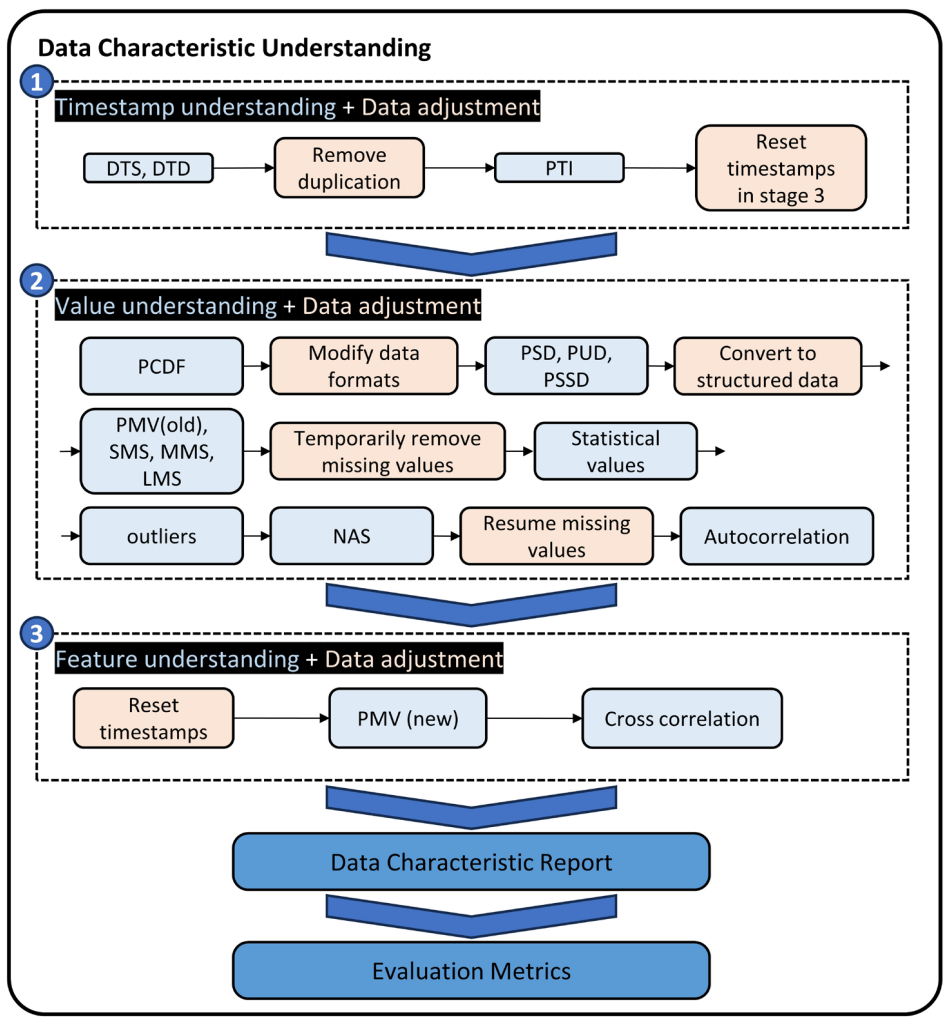

Figure 5. The pipeline of data characteristic understanding.

In the first stage - timestamp understanding, DTS and DTD are first calculated to find out duplicate timestamps. If the duplicate timestamps share the same value, it is easy to select one of them and remove other repeated records. If one timestamp matches different values, the most trustworthy value needs to be selected or computed based on the system property, expert recommendation, or statistical distributions. After removing duplications, PTI is computed to detect irregular time intervals. The standard time interval is a fixed value established by the system settings in the data acquisition stage. When consolidating all features into a dataset, irregular time intervals can lead to issues with value mapping. For instance, if it is assumed that values in the same row should correspond to the same time, discrepancies in time intervals among different features can result in values in the same row being recorded at different times. This discrepancy can subsequently disrupt the seamless processing of data. Therefore, the PTI computed in the stage will be utilized in the feature understanding to reset timestamps.

Following timestamp understanding, value understanding is employed. This begins with calculating PCDF to check the correctness of the data format, as an incorrect data format may lead to calculation issues. For instance, when a floating-point number is mistakenly recorded as an integer, it results in the loss of decimal information. After modifying the data formats, data forms are analyzed by computing PSD, PUD, and PSSD. The semi-structured and unstructured data should be converted into structured data as only structured data can be easily processed by most present machine learning algorithms and other data analytic approaches. Next, missing values are systematically analyzed by calculating PMV, SMS, MMS, and LMS. Missing values have a negative impact on the calculation of statistical values since they occupy a position in the dataset but do not contribute a value for computation. Although missing value imputation is a significant research topic in data preprocessing, it is not in the scope of this study, which aims to analyze data characteristics and make recommendations for data preprocessing to solve big data challenges. Therefore, this study employs a straightforward approach by simply removing the missing values. Then, the statistical values, outliers, and NAS are computed or detected to identify the characteristics regarding the statistical distribution of the time series. Subsequently, the missing values that were previously removed are reinstated for calculating autocorrelation. Autocorrelation relies on the time lag, and the removal of missing values disrupts the continuity of the time series.

The third stage is the feature characteristic understanding. To address the inconsistency of timestamps, irregular timestamps are detected by the PTI of all features in the first stage. New timestamps are established with proper intervals in the dataset. Values associated with irregular timestamps are then allocated to the nearest proper timestamp. In cases where multiple values need to be assigned to the same position, the appropriate one should be selected based on system characteristics or computed using statistical measures such as mean and median values. The new data instances lacking associated values are marked as missing values. Subsequently, the changes in data volumes result in an increased occurrence of missing values; therefore, the indicator PMV is re-calculated to monitor the current state of missing values. The adjustment of timestamps does not influence outliers and abnormal spikes, and they can be directly combined and utilized in this stage. Next, the computation of cross-correlations between features is performed to detect and analyze potential feature relationships.

After computing all characteristic indicators, evaluation metrics are calculated to demonstrate the overall quality of the dataset. These metrics provide valuable insights into the suitability of this pipeline for facilitating further decision-making of data preprocessing.

## 5. Case Study

In this section, the proposed framework is employed to understand data characteristics and evaluate big data challenges of two physical-sensor datasets originating from the manufacturing and transportation sectors respectively. The experiments are conducted in the Visual Studio Code platform (version: 1.85.2) and all codes are written in Python (version: 3.10.9). Section 5.1 provides detailed information about the datasets. Subsequently, the findings related to data characteristics are presented in Section 5.2.

## 5.1 Data description

### 5.1.1 Datasets in the manufacturing sector

The datasets are generated from a large Danish foundry, which is one of the largest foundries in Northern Europe. Figure 6 illustrates the foundry production process. The small light blue rectangles within the larger rectangle, labeled with the stages, represent the facilities associated with each stage. For instance, there are six induction furnaces depicted. The solid blue rectangles represent the movement of materials, and the circle with a cross denotes the completion of processing. Solid lines with arrows indicate the mandatory production flow, while dashed lines with arrows signify elective production flow, as ladle preheating may not be necessary for every production flow.

The production flow begins with collecting and sorting material from the scrap metal pits. Following this, iron is added to the induction furnaces and heated to a specified temperature, accounting for ferromagnetic losses. Once the melt reaches the desired temperature, it is transferred to a pre-heated transfer ladle. In this step, doping may be performed to impart specific alloy capabilities, aligning with the properties of the melt currently in the holding furnaces. The melt is then moved to a holding furnace, where it can be stored for a predetermined period. The transfer to the holding furnace signifies the completion of the melting phase of the foundry process. Subsequently, the primary forming stage commences to manufacture a variety of products.

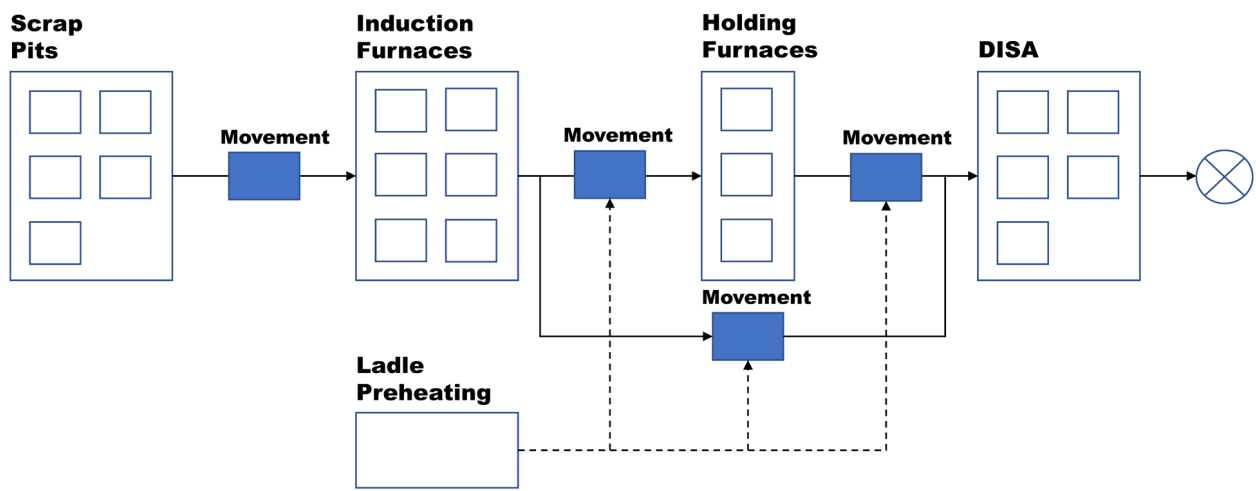

Figure 6. Industrial processing operations in the foundry.

The datasets used in this case study are generated by an IoT system, which includes a sensor network to monitor the operation conditions in the foundry production process. This study selects the data detecting the induction furnaces, the second stage in Figure 6, as they represent the most important step in the melting phase. In this case study, one of the six induction furnaces within the process is selected for analysis. It's noteworthy that all induction furnaces operate in parallel, and identical sensors are installed across all facilities. These sensors are responsible for detecting working temperatures and electricity consumption levels. The setting period for generating data is from December 21, 2022, to July 4, 2023 and the default time interval is 10 seconds. The seven features of the dataset are encoded to 'F1' - 'F7' respectively as Table 6 shows, which is designed to simplify and clarify the result presentation.

Table 6. Feature name encoding in the furnace dataset.

| Original name | Encoded name |
|---|---|
| TemperatureAct | F1 |
| PowerAct | F2 |
| CoolingFlowTemperature | F3 |
| CoolingReturnTemperature1 | F4 |
| CoolingReturnTemperature2 | F5 |
| CoolingReturnTemperature3 | F6 |
| CoolingReturnTemperature4 | F7 |

5.1.2 Datasets in the transportation sector

The datasets are acquired from the Danish public transportation system, and each represents the operational data of a bus on a specific route. The data are collected through both the embedded sensor system in the bus such as CANbus (Controller Area Network) and additional devices such as GPS and the acceleration meter. Figure 7 illustrates an example of a route, where the blue line refers to the route and the black hollow circles represent bus stops.

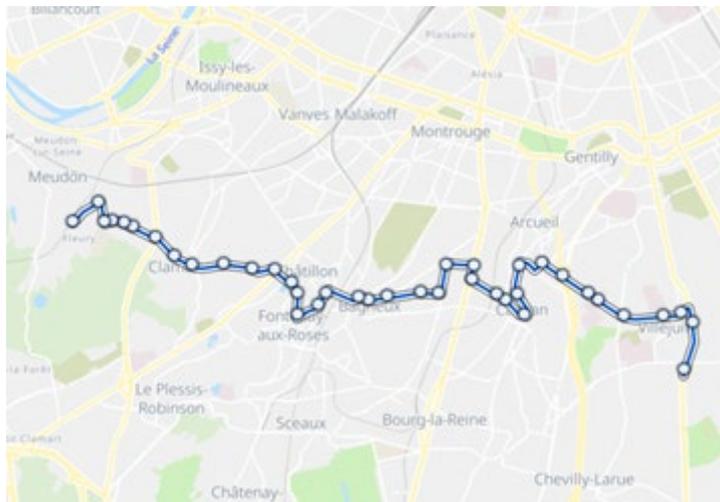

Figure 7. Example of a route.

The dataset in this case study involves one-day operational data with variables regarding speed, acceleration, and fuel consumption. The setting period of recording data is from 6:36:15 on August 7, 2023, to 21:12:12 on August 7, 2023 and the default time interval is 1 second. The ten features of the dataset are encoded to 'R1' - 'R10' respectively as Table 7 shows.

Table 7. Feature name encoding in the transportation dataset.

| Original name | Encoded name |
| --- | --- |
| TimeToLastMoment | R1 |
| Speed | R2 |
| CurrentAcceleration | R3 |
| ProcessedAcceleration | R4 |
| ConsumptionLastSecond | R5 |
| ComsumptionTotal | R6 |
| FuelRate | R7 |

| InstantFuelEco | R8 |
| TotalFuelUsed | R9 |
| TotalregenerativeEnergy | R10 |

## 5.2 Results of data characteristics understanding

### 5.2.1 Results of manufacturing data

This section shows the results of data characteristic understanding in the manufacturing case. Table 8 displays the evaluation scores of the 6Vs data characteristics quantified by the metrics defined in section 3.2. The findings reveal a large data volume, and good data variety but limited data velocity. The low score in data veracity underscores the commendable accuracy and reliability of the dataset. Moreover, the score of data value signifies a comprehensive data value representation, while the score of data variability indicates low data inconsistency.

Table 8. Evaluation scores of data characteristic understanding of the furnace dataset.

| 6Vs | Volume | Variety | Velocity | Veracity | Value | Variability |
|---|---|---|---|---|---|---|
| Evaluation metrics | Vol | Varie | Vel | Ver | Val | Varia |
| Scores | 11,317,010 | $+\infty$ | 10s | 1.128e-4 | 0.1571 | 0.1945 |

Table 9 shows the results of the data characteristic understanding of the furnace dataset, which is generalized based on the data characteristic indicators under 6Vs illustrated in Figure 3. Data volume includes indicators NF and NI, which represent the number of features and instances in the dataset. Results in the data variety show that all records in this dataset are structured data. Data velocity shows that the update speed is ten seconds because of the default settings of the IoT system.

Data veracity consists of metrics regarding data formats, time intervals, duplicate time, abnormal spikes, and missing values. The results show that all data formats in this dataset are correct, and 99.98% of time intervals in all features follow the default settings. In the case of all features, duplicate timestamps are associated with identical values, indicating that the data has been recorded erroneously on more than one occasion. 'F1' and 'F2' contain more abnormal spikes than other features, which might be the false temperature and power records due to sensor errors in high temperatures. The missing values are recorded in two datasets. The PMV (old) calculates the rate of missing values in the raw dataset, whereas the PMV (new) computes the rate of missing values in the adjusted dataset, whose timestamps are reset based on the pipeline introduced in Figure 5. It shows that the adjusted dataset contains more missing values because some timestamps are missed in the raw dataset. This is because certain sensors are closed in some specific operational conditions. Moreover, the majority of the missing time slices are short-term. In this metric, the short-term missing span is defined as 30-minute record missing, the medium-term missing span is 30-minute – 6-hour missing, and the long-term missing span is over 6-hour missing.

The metrics in the data value dimension are recorded in Table 10. The data validity dimension involves standard deviation in Table 10, cross correlation in Figure 8, and outlier rate. The results show that 'F1' has 1.282% outliers, 'F4' contains 0.01% outliers and others have no outliers. In the data preprocessing phase, it is imperative to replace these outlier values with appropriate and reasonable substitutes.

Table 9. Indicators of data characteristic understanding of the furnace dataset.

| Volume | NF | 7 | NI | 1,624,430 | | | | |
|---|---|---|---|---|---|---|---|---|
| Variety | PSD/PUD/PSSD | | F1 | F2 | F3 | F4 | F5 | F6 | F7 |
| | | | 100%/0/0 | 100%/0/0 | 100%/0/0 | 100%/0/0 | 100%/0/0 | 100%/0/0 | 100%/0/0 |
| Velocity | SDP | 10s | | | | | | | |
| Veracity | CDF | | F1 | F2 | F3 | F4 | F5 | F6 | F7 |
| | | | Yes | Yes | Yes | Yes | Yes | Yes | Yes |
| | PCDF | 100% | | | | | | | |
| | NAS | | F1 | F2 | F3 | F4 | F5 | F6 | F7 |
| | | | 117 | 120 | 1 | 1 | 0 | 0 | 0 |
| | PTI | | F1 | F2 | F3 | F4 | F5 | F6 | F7 |
| | | | 99.98% | 99.98% | 99.98% | 99.98% | 99.98% | 99.98% | 99.98% |
| | DTS/DTD | | F1 | F2 | F3 | F4 | F5 | F6 | F7 |
| | | | 162/0 | 162/0 | 162/0 | 162/0 | 162/0 | 162/0 | 162/0 |
| | PMV | | F1 | F2 | F3 | F4 | F5 | F6 | F7 |
| | old | | 0.044% | 0% | 0.023% | 0.023% | 0.023% | 0.023% | 0.023% |
| | new | | 3.505% | 3.465% | 3.465% | 3.465% | 3.465% | 3.465% | 3.465% |
| | SMS/MMS/LMS | | F1 | F2 | F3 | F4 | F5 | F6 | F7 |
| | | | 24/5/2 | 24/4/2 | 24/4/2 | 24/4/2 | 24/4/2 | 24/4/2 | 24/4/2 |
| Value | Statistical metrics | See Table 10 | | | | | | | |
| Variability | std | See Table 10 | | | cross correlation | | See Figure 8 | | |
| | outlier rate | | F1 | F2 | F3 | F4 | F5 | F6 | F7 |
| | | | 1.282% | 0% | 0% | 0.01% | 0% | 0% | 0% |

Table 10 shows the statistical values of each feature in the furnace dataset as the results of the data value. All these metrics show the distribution of each feature and the autocorrelation is illustrated in Figure 9. The minimum values for all features are zero because it is the default setting of the IoT system when the working mode is on standby. The seasonality of a univariate time series is detected using the 'seasonal_decompose' model in the Python library 'statesmodels 0.13.5' [78]. This model decomposes the time series into trend, seasonality, and noise in the additive form or the multiplicative form as Equations 21 and 22 show respectively.

$$x(t) = trend(t) + seasonality(t) + noise(t) \qquad (21)$$

$$x(t) = trend(t) \times seasonality(t) \times noise(t) \qquad (22)$$

where $x(t)$ represents the value of a time series $x$ at time $t$, $trend(t)$ refers to the value of the trend at time $t$, $seasonality(t)$ is the value of the trend at time $t$, and $noise(t)$ refers to the value of the noise at time $t$.

The additive form is selected in this case because the amplitude of the cycles is stable with time. The time series has no seasonality when the curve of the seasonality part is a line with constant values.

Table 10. Statistical values of each feature in the furnace dataset.

|  | F1 | F2 | F3 | F4 | F5 | F6 | F7 |
|---|---|---|---|---|---|---|---|
| cardinality | 1331 | 40902 | 246 | 682 | 571 | 579 | 605 |
| min | 0 | 0 | 0 | 0 | 0 | 0 | 0 |
| max | 62260.0 | 1200.0 | 37.4 | 95.2 | 80.8 | 79.4 | 80.9 |
| mean | 2238.45 | 320.59 | 24.3 | 29.08 | 32.34 | 31.85 | 32.12 |
| median | 1462.0 | 0.0 | 23.6 | 26.2 | 27.4 | 26.4 | 26.4 |
| std | 6837.38 | 461.15 | 5.28 | 11.68 | 14.55 | 14.18 | 14.79 |
| skewness | 8.66 | 0.82 | 0.03 | 1.13 | 0.63 | 0.64 | 0.63 |
| excess kurtosis | 70.05 | -4.22 | -4.1 | -2.23 | -4.17 | -4.19 | -4.24 |
| 1st qrt | 1411.0 | 0 | 19.2 | 19.3 | 19.5 | 19.6 | 19.3 |
| 3rd qrt | 1543.0 | 977.82 | 30.1 | 37.4 | 50.8 | 50.9 | 51.9 |
| seasonality | No | No | No | No | No | No | No |
| autocorrelation | See Figure 9 | | | | | | |

Figure 8 demonstrates the cross correlation of different features in the dataset, where the time delay is set at 300 seconds and the largest correlation value is selected as the result. It indicates that 'F1' has weak correlations with other features but others are strongly correlated, especially 'F5', 'F6', and 'F7'. This is because the operation temperature ('F1') changes frequently in the heating processing but the other features rely on the power settings.

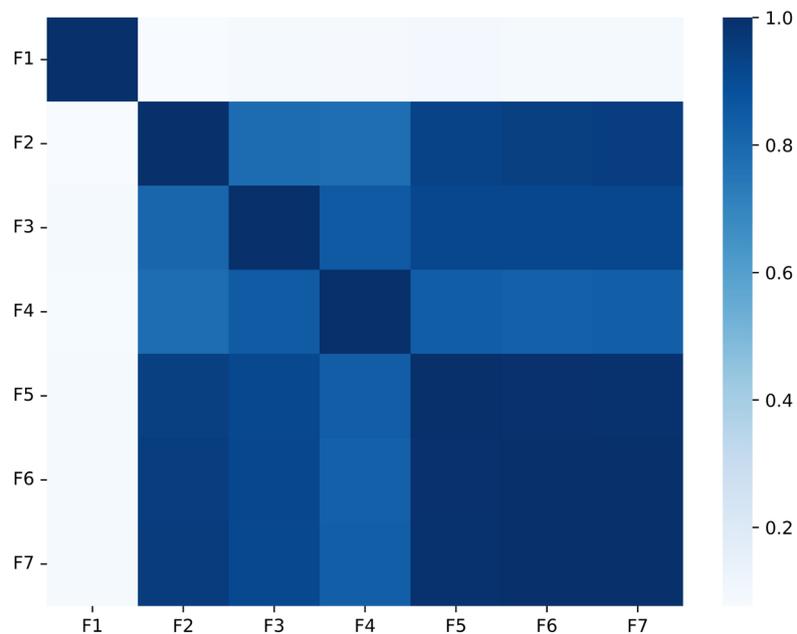

Figure 8. Cross correlation of different features in furnace dataset.

Figure 9 shows the autocorrelation results of each feature in the dataset. The lags in the figure denote the number of lagged values and a lag is 10 seconds due to the setting time interval. Hence, the maximum time lag is 600 seconds. The findings suggest that past operational activities have a significant impact on current operations within a 10-minute timeframe. However, this impact diminishes as time progresses.

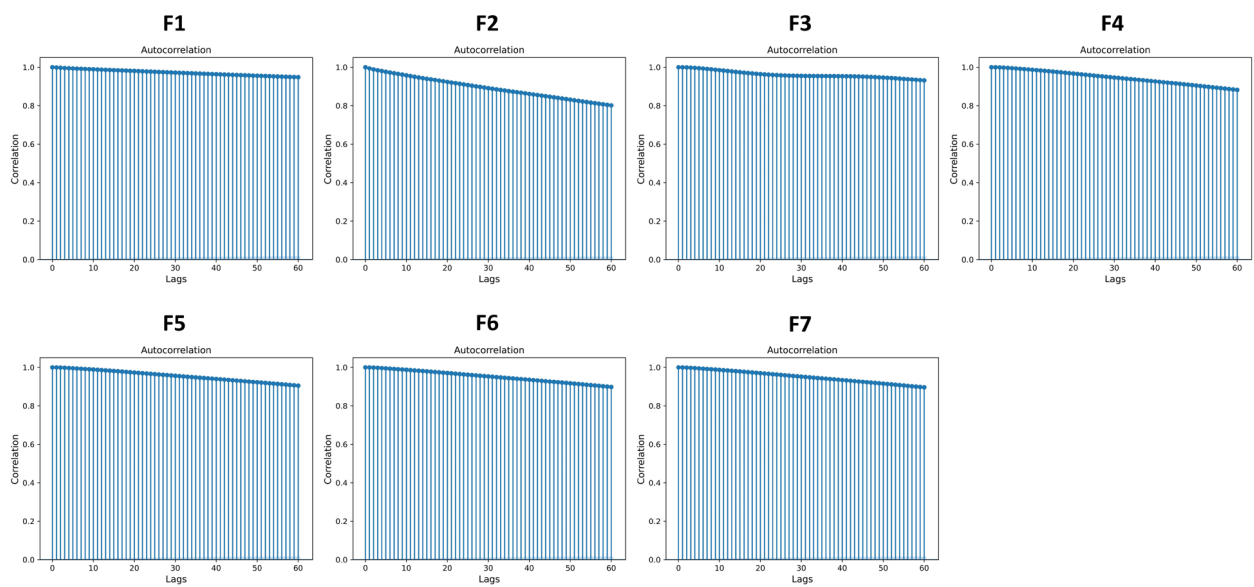

Figure 9. Autocorrelation of each feature in the furnace dataset.

5.2.2 Results of transportation data

Table 11 displays the evaluation scores of the 6Vs data characteristics of the transportation dataset. The findings reveal a large data volume, good data variety, and fast data velocity. The low score in data veracity suggests the commendable accuracy and reliability of the dataset. However, the score of data value signifies a less comprehensive representation of data value, while the score of data variability indicates high data inconsistency.

Table 11. Evaluation scores of data characteristic understanding of the transportation dataset.

| 6Vs | Volume | Variety | Velocity | Veracity | Value | Variability |
| --- | --- | --- | --- | --- | --- | --- |
| Evaluation metrics | Vol | Varie | Vel | Ver | Val | Varia |
| Scores | 435,230 | $+\infty$ | 1s | 3.818e-3 | 0.36 | 0.4229 |

Table 12 shows the results of data characteristic understanding of the transportation dataset and the structure is similar to that in Section 5.2.1. Data volume includes indicators NF and NI, representing the number of features and instances in the dataset. An analysis of data variety reveals that all records in this dataset are structured data. Data velocity shows that the update speed is 1 second because of the default settings of the IoT system monitoring the vehicle operations.

Data veracity consists of metrics regarding data formats, time intervals, duplicate time, abnormal spikes, and missing values. The results show that all data formats in this dataset are correct, and 99.99% of time intervals in all features follow the default settings. There is no duplicate timestamp in the dataset. The features 'R1', 'R3', and 'R4' contain several abnormal spikes. The abnormal spikes observed in 'R1' can be attributed to variations in recording gaps resulting from changes in vehicle operation modes. The irregular acceleration and braking patterns might result in the abnormal spikes detected in 'R3' and 'R4'. The missing values are recorded in two datasets as explained in Section 5.2.1. It shows that the adjusted dataset contains more missing values because some timestamps are missed in the raw dataset. This is because some specific sensors are deactivated when the vehicle motor is turned off or in standby mode. Moreover, the missing time slices in all features are long-term. In this metric, the short-term missing span is defined as 1-minute record missing, the medium-term missing span is 1-minute – 30-minute missing, and the long-term missing span is over 30-minute missing.

The metrics in the data value dimension are recorded in Table 13. The data validity dimension involves standard deviation in Table 13, cross correlation in Figure 10, and outlier rate. The results show that the features 'R1', 'R3', 'R4', and 'R5' contain outliers and others have no outliers. In the data preprocessing phase, it is imperative to replace these outlier values with appropriate and reasonable substitutes.

Table 12. Data characteristic understanding of the transportation dataset.

| Volume | NF | 10 | | | NI | 43523 | | | | | |
|---|---|---|---|---|---|---|---|---|---|---|---|
| Variety | PSD/PUD/PSSD | R1 | R2 | R3 | R4 | R5 | R6 | R7 | R8 | R9 | R10 |
| | | 100%/0/0 | 100%/0/0 | 100%/0/0 | 100%/0/0 | 100%/0/0 | 100%/0/0 | 100%/0/0 | 100%/0/0 | 100%/0/0 | 100%/0/0 |
| Velocity | SDP | 1s | | | | | | | | | |
| Veracity | CDF | R1 | R2 | R3 | R4 | R5 | R6 | R7 | R8 | R9 | R10 |
| | | Yes | Yes | Yes | Yes | Yes | Yes | Yes | Yes | Yes | Yes |
| | PCDF | 100% | | | | | | | | | |
| | NAS | R1 | R2 | R3 | R4 | R5 | R6 | R7 | R8 | R9 | R10 |
| | | 62 | 0 | 347 | 245 | 0 | 0 | 0 | 0 | 0 | 0 |
| | PTI | R1 | R2 | R3 | R4 | R5 | R6 | R7 | R8 | R9 | R10 |

|  |  |  |  |  |  |  |  |  |  |  |
|---|---|---|---|---|---|---|---|---|---|---|
|  |  | 99.99% | 99.99% | 99.99% | 99.99% | 99.99% | 99.99% | 99.99% | 99.99% | 99.99% |
|  | DTS/DTD | R1 | R2 | R3 | R4 | R5 | R6 | R7 | R8 | R9 | R10 |
|  |  | 0/0 | 0/0 | 0/0 | 0/0 | 0/0 | 0/0 | 0/0 | 0/0 | 0/0 | 0/0 |
|  | PMV | R1 | R2 | R3 | R4 | R5 | R6 | R7 | R8 | R9 | R10 |
|  | old | 0% | 0% | 13.60% | 0.039% | 0.005% | 0.005% | 0.005% | 0.005% | 0.005% | 0.005% |
|  | new | 17.15% | 17.15% | 28.42% | 17.18% | 17.15% | 17.15% | 17.15% | 17.15% | 17.15% | 17.15% |
|  | SMS/MMS/LMS | R1 | R2 | R3 | R4 | R5 | R6 | R7 | R8 | R9 | R10 |
|  |  | 0/0/1 | 0/0/1 | 388/0/2 | 1/0/1 | 1/0/1 | 1/0/1 | 1/0/1 | 1/0/1 | 1/0/1 | 1/0/1 |
| Value | Statistical metrics | See Table 13 |
| Variability | std | See Table 13 |  | Cross Correlation |  | See Figure 10 |
|  | Outlier rate | R1 | R2 | R3 | R4 | R5 | R6 | R7 | R8 | R9 | R10 |
|  |  | 2.521% | 0% | 54.67% | 52.73% | 0.466% | 0% | 0% | 0% | 0% | 0% |

Table 13 presents the statistical values of each feature in the transportation dataset, serving as the outcomes of the data value. These metrics provide insights into the distribution of individual features, while autocorrelation is visually depicted in Figure 11. The additive form is selected in this case as the amplitude of the cycles is stable with time. Notably, the time series exhibits no seasonality, as evidenced by a seasonality part represented by a constant value line in the curve.

Table 13. Statistical values of each feature in the transportation dataset.

|  | R1 | R2 | R3 | R4 | R5 | R6 | R7 | R8 | R9 | R10 |
|---|---|---|---|---|---|---|---|---|---|---|
| cardinality | 52 | 7579 | 29290 | 30036 | 2 | 201 | 1 | 1 | 201 | 1 |
| min | 0 | 0 | -3.0 | -8.44 | 0 | 374940.5 | 0 | 0 | 749881.0 | 0 |
| max | 1066 | 87.83 | 4.62 | 19.49 | 0.5 | 375040.5 | 0 | 0 | 750081.0 | 0 |
| mean | 1000.59 | 22.5 | -0.02 | -0.07 | 0 | 374998.5 | 0 | 0 | 749996.99 | 0 |
| median | 1001.0 | 12.77 | 0 | 0 | 0 | 374996.5 | 0 | 0 | 749993.0 | 0 |
| std | 6.99 | 25.04 | 0.43 | 0.5 | 0.03 | 31.23 | 0 | 0 | 62.46 | 0 |
| skewness | -134.67 | 0.76 | -0.55 | 0.66 | 14.54 | -0.09 | N/A | N/A | -0.09 | N/A |
| excess kurtosis | 19275.84 | -3.69 | 3.14 | 59.16 | 206.39 | -4.3 | N/A | N/A | -4.3 | N/A |
| 1st qrt | 1000 | 0.03 | -0.17 | -0.24 | 0 | 374965.5 | 0 | 0 | 749931.0 | 0 |
| 3rd qrt | 1001 | 46.17 | 0.16 | 0.14 | 0 | 375034.0 | 0 | 0 | 750068.0 | 0 |
| seasonality | No | No | No | No | No | No | No | No | No | No |
| autocorrelation | See Figure 11 |

Figure 10 illustrates the cross correlation of various features in the dataset, with a time delay set at 300 seconds, and the largest correlation value is selected as the result. It indicates a strong correlation between features 'R3' and 'R4', as well as between features 'R6' and 'R9'. 'R3' and 'R4' represent two different measurements of acceleration, while 'R6' and 'R9' are two different detection of fuel consumption.

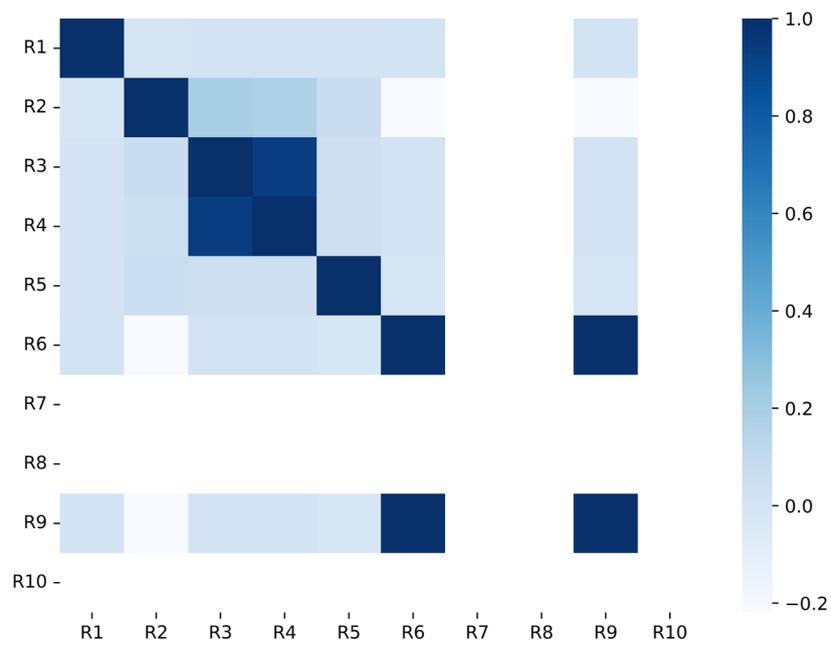

Figure 10. Cross correlation of different features in the transportation dataset.

Figure 11 illustrates the autocorrelation results of each feature in the dataset. The lags in the figure denote the number of lagged values and a lag is 10 seconds due to the setting time interval. Hence, the maximum time lag is 600 seconds. The present time series recorded in features 'R1', 'R5', 'R7', 'R8', and 'R10' appear to be unaffected by their historical data, suggesting that the trend of historical data is not suitable for technical analysis. Additionally, the autocorrelation coefficients of the data in features 'R3' and 'R4' rapidly diminish from high values, indicating a strong correlation within a 20-second timeframe. The data in features 'R2', 'R6', and 'R9' suggest that past vehicle operations have a significant impact on current modes within a 10-minute timeframe although this impact diminishes as time progresses.

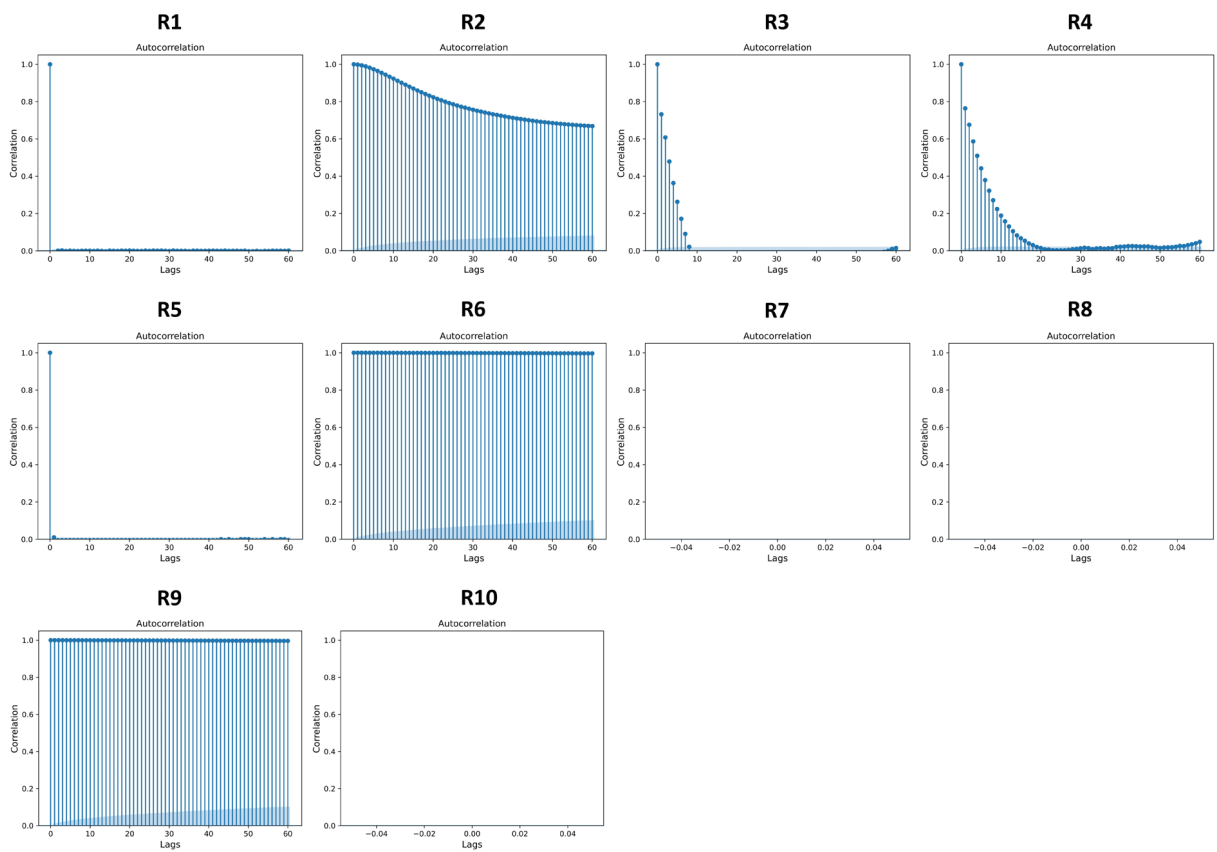

Figure 11. Autocorrelation of each feature in the transportation dataset.

## 6. Discussions

In this section, firstly the big data challenges in the furnace dataset and transportation dataset are discussed based on the results in Section 5, followed by specific preprocessing recommendations tailored to each dataset. Furthermore, the proposed data characteristic understanding model with existing models is compared, followed by highlighting the advantages of incorporating selected dimensions and providing statistical indicators for a more comprehensive analysis.

## 6.1 Data Challenge Analysis and Data Preprocessing Recommendation

In this subsection, an in-depth analysis of the big data challenges for each dataset is conducted, and corresponding data preprocessing recommendations are presented, drawing from the discussions outlined in Section 3.3. The assessment of big data challenges relies on the outcomes of data characteristic understanding as detailed in Section 6.1.2, guiding the formulation of preprocessing recommendations tailored to address the identified challenges. Because the challenges in the data volume and value dimensions need to be discussed alongside other dimensions as explained in Sections 3.3.1 and 3.3.5, this section exclusively addresses the remaining four dimensions.

### 6.1.1 Data challenges and preprocessing in the furnace dataset

Firstly, all data points in the furnace dataset are structured data, eliminating challenges in the data variety dimension. However, a notable concern arises with the data being updated every 10 seconds, posing a challenge in the data velocity dimension as this time interval is relatively long, leading to the loss of valuable information, such as mode changes. It is recommended to increase the frequency of data recording to 1 second for each data point. This adjustment aims to capture more detailed information and mitigate the loss of valuable data.

The challenges in the data variety are derived from the indicators regarding data quality. The data formats are all correct and no corrections are needed. Abnormal spikes can be addressed by replacing the values through linear interpolation. This method is easily implemented although it may have lower accuracy compared to machine learning algorithms. However, given that the abnormal spikes constitute less than 0.01% in four of the seven features, the impact on the overall dataset analytics is minimal, making this a pragmatic and efficient solution. Next, the inconsistent time intervals have been addressed in the data characteristic understanding pipeline in Section 4. Furthermore, since all duplicate timestamps have identical values, they can be seamlessly integrated into a single data instance by removing the redundant records. Finally, the missing values are the largest challenge in this dataset. The short-term and medium-term missing spans can be imputed through some AI algorithms. However, the long-term missing spans cannot be reliably imputed due to the absence of seasonality in the time series. Therefore, the recommended approach is to select a subset that excludes the segments with long-term missing spans.

The challenge of data inconsistency in the data variability dimension consists of the outliers and the cross-correlation. The features 'F1' and 'F4' contain outliers and the processing decision should be made in collaboration with the maximum and minimum values provided in Table 9. The range of temperatures in feature 'F4' spans from 0°C to 95.2°C, which falls in a reasonable and expected range of the water in the cooling loop. However, the maximum action temperature is 62260°C, which is much higher than the default setting (1460°C). In this case, it is essential to examine the operation mode, and reasonable values can be estimated based on both the operation modes and the surrounding values. Furthermore, following the cross-correlation analysis, it is evident that no features demonstrate a strong relationship with the feature 'F1.' However, it is crucial to highlight the related features with the action temperature in the context of this industrial processing. Therefore, some other technologies such as feature engineering and causal inference should be applied to identify the feature relationships.

6.1.2 Data challenges and preprocessing in the transportation dataset

In the transportation dataset, all data points are characterized by structured data, thereby mitigating challenges associated with data variety. Furthermore, the time interval for data collection is set at 1 second, ensuring a rapid update rate that facilitates the timely gathering of information.

The challenges related to data variety stem from the indicators assessing data quality. The data formats are all correct, requiring no corrections. Abnormal spikes can be addressed by replacing the values through linear interpolation because they constitute less than 1% in three of the ten features. Next, the inconsistent time intervals have been addressed in the data characteristic understanding pipeline in Section 4. Furthermore, there are no duplicate timestamps in the dataset. Finally, the missing values should be addressed in this dataset. As explained in Section 6.1, the short-term missing spans can be effectively imputed using various AI algorithms, and a subset can be chosen excluding segments that encompass the long-term missing spans.

The challenge of data inconsistency in the data variability dimension encompasses the outliers and the cross-correlation. The features 'R1', 'R3', 'R4', and 'R5' contain outliers and the processing decision should be made in collaboration with the maximum and minimum values provided in Table 11. The values in these four features fall within reasonable ranges, negating the need for outlier processing as all values fall within valid ranges. Furthermore, the cross-correlation analysis reveals no meaningful results regarding features 'R7', 'R8', and 'R10' as their values constantly remain zero. Therefore, it is advisable to exclude these three features from further analytics due to their inability to provide meaningful information.

## 6.2 Comparison with existing models

The results of the proposed data characteristic understanding model are compared to seven existing models to verify the contributions of this paper. The evaluation and comparison of data characteristic understanding models encompass five dimensions: the number of Vs dimensions (Dim.), the number of indicators (Indi.), quantitative ability (Quanti.), capacity for processing panel data (Panel), and capability for processing time series (Time). The results are presented in Table 14. In Table 14, 'n/a' denotes no applicable data, while 'Y' and 'N' represent yes and no, respectively.

Table 14. Model comparison under five dimensions.

| Ref. | Dim. | Indi. | Quanti. | Panel | Time |
|---|---|---|---|---|---|
| [7] | 4 | 4 | Y | Y | N |
| [8] | 6 | 13 | Y | Y | Y |
| [9] | 5 | n/a | N | Y | Y |
| [12] | 6 | n/a | N | Y | N |
| [13] | 6 | n/a | N | Y | N |
| [14] | 6 | n/a | N | Y | N |
| [66] | n/a | 26 | Y | N | Y |

| Proposed Framework | 6 | 30 | Y | Y | Y |

Based on the comparison in Table 14, the three limitations of the existing models introduced in Section 2.2 have been addressed by the proposed framework. Firstly, the selected dimensions of the proposed 6Vs model have mutual and unique definitions and there is no overlap among them.

The proposed framework exhibits the highest number of dimensions and the most clearly defined mutual understanding of these dimensions compared to the other seven models. Moreover, it encompasses the largest array of indicators, thereby effectively quantifying the Vs dimensions. For example, The data quality evaluation framework developed in [7] employs volume, velocity, variety, and veracity to assess data quality. However, the model presented in this article incorporates two additional dimensions, namely value and variability, offering more comprehensive and valuable information. The model in [8] introduces vincularity in its 6Vs model and quantifies this dimension using an indicator named traceability. However, the popularity of this dimension is limited in the literature discussed in Section 2.2, rendering its definition less convincing. In contrast, the dimension 'Variability' enjoys widespread usage, with its definition being consistently mutual in many scholarly works.

Secondly, the existing 6Vs models in [9, 12-14] only provide qualitative definitions and analyses, lacking quantitative metrics. This limitation restricts a thorough assessment of data characteristics. In contrast, the proposed framework assigns statistical indicators to each dimension and unveils the big data challenges. Additionally, data preprocessing recommendations are derived from these indicators to improve the data quality.

Lastly, no data characteristic understanding models are developed specifically for the physical-sensor datasets in the literature study. Many models [7, 9, 12-14] can be employed to analyze all types of big data and do not involve time-related characteristic indicators. The study in [66] analyzes diverse and controllable time-series characteristics for generating new high-quality time series. However, this framework lacks a systematic analysis of data characteristics and does not examine timestamps. The proposed framework incorporates a comprehensive set of characteristic indicators for physical-sensor data. This includes indicators for panel data values, time series, and timestamps.

## 6.3 Model generalizability

The proposed framework is developed for understanding physical-sensor data characteristics, which has the potential to be applied to explore other types of datasets and in other domains. The indicators in the framework elucidate the characteristics of panel data values, time series and timestamps, which are the general properties not only in the physical-sensor data. Some of these indicators can be employed to understand other types of data. For instance, survey data typically exhibit panel data attributes, encompassing both continuous and categorical features. The indicators aimed at characterizing these features, irrespective of temporal considerations, can be employed in such scenarios. Additionally, financial data are usually time series, and the indicators detecting the temporal trends, seasonality, autocorrelation and so on can provide valuable insights in such scenarios. The flexibility of the framework lies in its ability to accommodate different data formats by leveraging various techniques such as statistical modeling, data visualization, time series analysis and so on. Through systematic adaptation and refinement, it shows promise for advancing understanding and analysis beyond physical sensor data, contributing to a broader understanding of big data characteristics.

## 7. Conclusion

The characteristics of big data have captured the interest of researchers in the context of physical-sensor big data. These data characteristics provide high-quality insights and recommendations for further data analytics, enabling data-driven decisions to effectively tackle various data challenges. This article proposes a systematic model for understanding data characteristics, aiming to analyze the quantitative aspects of physical-sensor data characteristics and the associated data challenges. The insights gained from understanding data characteristics can be employed to aid in the data preprocessing.

The data characteristics understanding is based on a 6Vs model, including volume, variety, velocity, veracity, value, and variability. Each dimension incorporates a set of statistical indicators, enhancing the objectivity of the model for understanding data characteristics. Moreover, several indicators are associated with the temporal data characteristics, given that physical-sensor data are inherently time-related. Furthermore, a big-data challenge is linked to each data characteristic dimension in the 6Vs model to comprehend the challenges posed by the physical-sensor data. The potential data preprocessing strategies are recommended to effectively address these challenges. Finally, a pipeline to implement the proposed framework for understanding the characteristics of the physical-sensor data is developed in this article, and the pipeline encompasses timestamp understanding, value understanding, and feature understanding. Two case studies are conducted to visualize the utilization of the proposed framework. The data are generated from the industrial processing and transportation sectors respectively, and both datasets are collected by the physical sensors. The results reveal the data characteristics of each dataset in six dimensions. Data challenges are then derived from the characteristic indicators, and recommendations for data preprocessing to address these challenges are also discussed.

Data characteristics and challenging understanding constitute the initial stages of the CRISP-DM-based data lifecycle, aiming to provide precise and accurate information to enhance data quality for subsequent data analytics. In future studies, a data quality evaluation framework should be developed for quantitative data quality profiling and assessment. Furthermore, a comprehensive data preprocessing framework should be developed based on the results of data characteristics understanding and data quality assessment.

# List of abbreviations

| | |
|---|---|
| IoT: | Internet of Things |
| DLC: | Data Lifecycle |
| CRISP-DM: | The Cross Industry Standard Process for Data Mining |
| APREP-DM: | The Automated Preprocessing for Data Mining |
| NF: | The Number of Features |
| NI: | The Number of Instances |
| PSD: | The Percentage of Structured Data |
| PUD: | The Percentage of Unstructured Data |
| PSSD: | The Percentage of Semi-structured Data |
| SDP: | The Speed of Data Processing |
| CDF: | The Correctness of Data Formats |
| PCDF: | The Percentage of the Correctness of Data Formats |
| NAS: | The Number of Abnormal Spikes |
| PTI: | The Percentage of Normal Time Intervals |
| DTS: | Duplicate timestamps with the same value |
| DTD: | Duplicate timestamps with different values |
| MCAR: | Missing Completely at Random |
| MAR: | Missing at Random |
| MNAR: | Missing not at Random |
| PMV: | The Percentage of Missing Values for a Feature |
| SMS: | Short-term Missing Span |
| MMS: | Medium-term Missing Span |
| LMS: | Long-term missing Span |

# Declarations

### Ethics approval and consent to participate

No appliable.

### Consent for publication

No appliable.

### Availability of data and materials

The datasets used and analyzed during the current study are available from the corresponding author on reasonable request.

### Competing interests

The authors declare that they have no competing interests.

### Funding

This paper is part of the project "IEA IETS Task XVIII: Digitalization, Artificial Intelligence and Related Technologies for Energy Efficiency and GHG Emissions Reduction in Industry", funded by the Danish funding agency, the Danish Energy Technology Development and Demonstration (EUPD) program, Denmark (Case no.134-21010), the project "Driver Coach", funded by EUDP (Case no.64021-2034) and the project "Data-driven best-practice for energy-efficient operation of industrial processes - A system integration approach to reduce the CO2 emissions of industrial processes" funded by EUDP (Case no.64020-2108).

## Authors' contributions

ZM contributes to the conception, methodology, framework design, data acquisition and analysis, programming, writing the draft manuscript and revisions. BNJ works on the conception, methodology, framework improvement, manuscript revisions and supervision. ZGM works on the conception, methodology, framework improvement, manuscript revisions and supervision. All authors have read and approved the final manuscript.

## Acknowledgements

No appliable.